\documentclass{elsart}

\usepackage{graphicx}
\usepackage{algorithm}
\usepackage{algorithmic}
\usepackage{amsmath}
\usepackage{url}

\begin{document}

\begin{frontmatter}
  
  \title{Improved neighbor list algorithm in molecular simulations
    using cell decomposition and data sorting method}
  
  \author[yao]{Zhenhua Yao\corauthref{cor-yao}},
  \ead{smayzh@nus.edu.sg}
  \author[yao,wang]{Jian-Sheng Wang},
  \author[liu]{Gui-Rong Liu}
  \author[cm]{Min Cheng}
  
  \address[yao]{The Singapore-MIT Alliance, National University of
    Singapore, Singapore 117576}
  
  \address[wang]{Department of Computational Science, National
    University of Singapore, Singapore 117543}
  
  \address[liu]{Centre for Advanced Computations in Engineering
    Science (ACES), Department of Mechanical Engineering, National
    University of Singapore, Singapore 119260}

  \address[cm]{Department of Communication Engineering, Nanyang
    Technological University, Singapore 639798}

  \corauth[cor-yao]{Corresponding author.}

  \begin{abstract}
    An improved neighbor list algorithm is proposed to reduce
    unnecessary interatomic distance calculations in molecular
    simulations. It combines the advantages of Verlet table and
    cell linked list algorithms by using cell decomposition approach
    to accelerate the neighbor list construction speed, and data
    sorting method to lower the CPU data cache miss rate, as well as
    partial updating method to minimize the unnecessary reconstruction
    of the neighbor list.  Both serial and parallel performance of
    molecular dynamics simulation are evaluated using the proposed
    algorithm and compared with those using conventional Verlet table
    and cell linked list algorithms.  Results show that the new
    algorithm outperforms the conventional algorithms by a factor of
    $2\sim 3$ in cases of both small and large number of atoms.
  \end{abstract}

  \begin{keyword}
    Molecular dynamics \sep Neighbor list \sep Verlet table \sep
    Cell linked list
    \PACS 02.70.Ns \sep 31.15.Qg \sep 33.15.Dj \sep 87.15.He
  \end{keyword}

  \journal{Computer Physics Communications}

\end{frontmatter}


\section{Introduction}
\label{sec:intro}

Some molecular simulation techniques such as molecular dynamics and
Monte Carlo method are widely used to study the physical properties
and chemical processes which contain a large number of atoms in
statistical physics, computational chemistry, and molecular biology
\cite{thijssen}. All these methods involve evaluation of the total
interatomic potential energy $V_{\text{tot}}$ of $N$ atoms and its
gradients. The potential energy contains various interatomic
interactions in the physical system, and is usually the function of
internal coordinates of atoms. For example the potential energy of
liquids and gases is often described as a sum of two-body (or
pairwise) interactions over all atom pairs. A common choice of
two-body interatomic interaction expression is Lennard-Jones potential
function, which is a simple function of the distance $r_{ij}$ between
atom $i$ and $j$, and is shown as follows,
\begin{equation}
  \label{eq:lj}
  V_{\text{LJ}} (r_{ij}) = 4\epsilon\left[ \left(\frac{\sigma}{r_{ij}}
      \right)^{12} -  \left(\frac{\sigma}{r_{ij}} \right)^6 \right].
\end{equation}
The total potential energy is a sum of two-body interactions over all
atom pairs,
\begin{equation}
  \label{eq:tot-pe}
  V_{\text{tot}} = \frac{1}{2} \sum_{i=1}^{N} \sum_{\substack{j=1\\j\neq i}}^N
  V_{\text{LJ}}(r_{ij}).
\end{equation}

In molecular dynamics simulation, evaluation of Eq.\eqref{eq:tot-pe}
and its gradient usually costs most of CPU time. Apparently, direct
calculation of Eq.\eqref{eq:tot-pe} requires $N^2$ steps. If we use
Newton's third law, the total calculation steps can be decreased to
$N(N-1)/2$. Obviously it is formidable to carry out such a calculation
when there are many atoms in the system, and some methods are strongly
needed to reduce the redundancy in evaluation of Eq.\eqref{eq:tot-pe}.

Firstly a cutoff distance $r_{\text{cut}}$ is introduced in potential
functions, and both potential functions and their gradients beyond the
cutoff distance are assumed to be zero. This treatment can reduce the
computing time greatly by neglecting all atoms beyond the cutoff
distance, since interactions between these atoms are zero and needn't
to be considered. However, straightforward determination of which
atoms are within cutoff distance needs to evaluate all interatomic
distances over all atom pairs, and this procedure scales $O(N^2)$ as
the system size.

Effective reduction of unnecessary interatomic distance evaluation can
be accomplished by Verlet table algorithm \cite{verlet} and cell
linked list algorithm \cite{cell}. However, there is a tradeoff
between overhead for maintaining neighbor list table and reduction of
unnecessary interatomic distance calculation.

Verlet table algorithm and cell linked list algorithm have been
intensively studied and have shown significant reduction in total
computing time \cite{mattson99,walther01,matin,weiser,ariel,rycerz}.
Glikman \textit{et al} proposed an alternative Verlet table method
called \emph{relational method} which updates the neighbor list
recursively and is called in every timestep \cite{glikman}, Eisenhauer
considered the locality of the neighbor list and used a simple
slab-based decomposition in his work \cite{greg}. In 1994, Hansen and
Evans employed cell method in conventional Verlet table algorithm to
accelerate the neighbor list construction speed \cite{hansen}, and
Frenkel also considered combination of two conventional neighbor list
algorithms in his book \cite{frenkel}. In our work improved neighbor
list algorithm is proposed and the overhead to maintain the neighbor
list table has been reduced to order $O(N)$.  The details and
benchmark of this algorithm are given in the paper.


\section{Conventional neighbor list algorithms and the improved
  algorithm}
\label{sec:method}

For the details of conventional Verlet table and cell linked list
algorithms one can refer to Allen and Tildesley's book \cite{allen},
and here we only mention some fundamental and important facts.  Some
graphs in this section are drawn in 2D for the convenience of
illustration, however, all related discussions can be easily
generalized to 3D systems.

\subsection{Conventional Verlet table algorithm}
\label{sec:vt-method}

The basic idea of Verlet table algorithm is to construct and maintain
a list of neighboring atoms for each atom in the system
\cite{verlet,allen}.  During the simulation, the neighbor list will be
updated periodically in a fixed interval, or automatically when the
displacements of some atoms are larger than a certain value
\cite{fincham}. In this algorithm the potential cutoff sphere with
radius $r_{\text{cut}}$ is surrounded by a ``skin'' $r_s$, to give a
larger sphere with radius $r_{\text{s}} + r_s$ \cite{allen}. When
constructing a neighbor list for an atom $i$, another atom $j$ is
considered as a ``neighbor'' if the distance between them $r_{ij} \le
r_{\text{cut}}+r_s$.  It should be noticed that the ``skin'' $r_s$
should be large enough, so that between two times of neighbor list
reconstruction no atom can penetrate through the skin into the cutoff
sphere of another atom if it is not in the neighbor list of that atom.
\begin{figure}[hbtp!]
  \centering
  \includegraphics[width=0.3\textwidth]{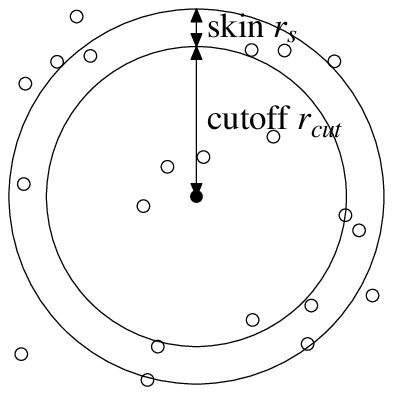}
  \caption{Illustration of conventional Verlet table algorithm}
  \label{fig:verlet}
\end{figure}


Conventionally in this algorithm, interatomic distances of all atom
pairs need to be calculated, so the total CPU time scales $O(N^2)$ as
the number of atoms. However, evaluation of the total potential energy
and its gradient using the neighbor list is efficient because only
atoms appearing in the list, i.e., those in the sphere of
$r_{\text{cut}}+r_s$, are checked, thus the overall procedure scales
$O(N \cdot N_{\text{NL}})$, where $N_{\text{NL}}$ is the average
number of neighbors of an atom in the material and is independent to
the system size $N$.

Verlet table algorithm has been proven to be efficient when the number
of atoms is relatively small and the atoms move slowly.  Its main
advantage is the high efficiency of potential/force evaluation using
the neighbor list, as the atoms in the list are only a few more than
actual needed (which $r_{\text{cut}}<r_{ij}<r_{\text{cut}}+r_s$).  On
the other hand, its main drawback is that construction of the neighbor
list scales $O(N^2)$ (needs $N(N-1)/2$ steps to build the list for all
atoms).  Moreover, with the increase of atoms' mobility, either the
``skin'' or the frequency of reconstructing the neighbor list must
increase.  Both of them make the overall simulation time increases
dramatically.

\subsection{Conventional cell linked list algorithm}
\label{sec:ll-method}

The cell linked list algorithm is effective when the number of atoms
is large. In this algorithm, the simulation domain is partitioned into
several cells, and the edge of cells is equal to or larger than
$r_{\text{cut}}$. All atoms are assigned to these cells by their
positions, at the same time a linked list of the atom indices is
created. At the beginning of a simulation, an array containing a list
of cell neighbors for each cell is created, and this list remains
fixed unless the simulation domain changes during the simulation
\cite{allen}.

A cell $m$ is considered as a neighbor of another cell $n$, if there
are a point at $\mathbf{r}_i$ in cell $m$ and another point at
$\mathbf{r}_j$ in cell $n$ such that $|\mathbf{r}_i-\mathbf{r}_j| \le
r_{\text{cut}}$.  Since in the conventional method the edge of the
cell is equal to or larger than $r_{\text{cut}}$, and considering the
periodic boundary condition, there are 8 (for 2D systems) or 26 (for
3D systems) neighbors for each cell. Thus the neighbors of each atom
can be found in the cell where the atom located and the neighboring
cells.

The construction of ``neighbor list'', i.e., assigning each atom to
appropriate cells, scales $O(N)$, but we can see that a big number of
atoms need to be checked in the potential/force evaluation procedure,
and this is rather inefficient compared to Verlet table algorithm.  A
common choice of the cell edge is the potential cutoff distance
$r_{\text{cut}}$, thus for each atom, all atoms in 27 cells, or in the
volume of $27 r_{\text{cut}}^3$, will be checked in the
potential/force evaluation procedure\footnote{If Newton's 3rd law is
  used and the interatomic interaction is described by a two-body
  potential function, only half of neighboring cells needs to be
  searched, in other words, only 14 cells are needed for 3D systems.
  However, if multiple-body potential functions are adopted,
  especially those using bond order terms, we cannot safely do this.
  For example in Brenner potential, calculation of $N_{ij}^{conj} = 1
  + [ \sum_{k \ne i,j}^{carbon}f_{ij}^c(r_{ik}) F(X_{ik}) ]^2 + [
  \sum_{l \ne i,j}^{carbon}f_{jl}^c(r_{jl}) F(X_{jl}) ]^2$, i.e.,
  determination of whether a bond is part of a conjugated system or not,
  needs to reference all neighbors of every atom. In this case,
  searching only half of the neighboring cells is not correct.
  Therefore, in general cases we say all 27 cells need to be
  searched.}. Ideally, only atoms in the volume of $\frac{4}{3}\pi
r_{\text{cut}}^3 \approx 4.189 \; r_{\text{cut}}^3$ fall in the cutoff
distance, and accordingly need to be checked.

\begin{figure}[hbtp!]
  \centering
  \includegraphics[width=0.45\textwidth]{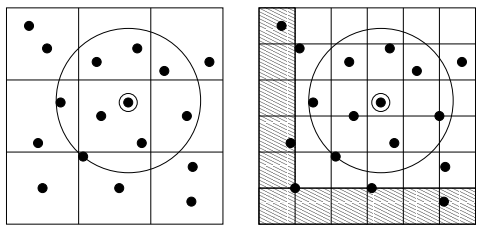}
  \caption{Choice of the length of cell edge in the cell linked list
    algorithm: $r_{\text{cut}}$ vs $\frac{1}{2}r_{\text{cut}}$.}
  \label{fig:cell}
\end{figure}

However, if a small cell edge is used, volume containing the atoms
need to be checked will be dramatically reduced. For example if the cell
edge is $\frac{1}{2} r_{\text{cut}}$, the volume will be $(2.5)^3 \;
r_{\text{cut}}^3$, only 57.87\% of the previous volume, as
Fig.~\ref{fig:cell} shows for 2D systems.

Furthermore, the cell edge can be very small so that only one atom can
be contained in the cell, as described in Ref.~\cite{allen}. This
method is reported to has higher performance than conventional
algorithm. However, according to our tests, it is still not as fast as
Verlet table algorithm when the number of atoms is relatively small.

From above discussions, we can see that the advantage of the cell
linked list algorithm is the fast and efficient construction of the
``neighbor list'', but its disadvantage is that too many atoms need to
be checked in the potential/force evaluation procedure, and
improvements to this algorithm are not competitive to Verlet table
algorithm with the increasing of implementation complexity when the
number of atoms is relatively small.

\subsection{The improved neighbor list algorithm}
\label{sec:bvt}


Verlet table algorithm will scale linearly as the number of atoms if
cell decomposition approach is adopted. Together with data sorting and
partial updating method, we get an improved neighbor list algorithm
with high performance.

\subsubsection{Cell decomposition approach}
\label{sec:cell decomp}

As cell decomposition approach has been introduced and used previously
\cite{frenkel,hansen}, here we only give its main features.  In this
approach two conventional algorithms are combined together.  Firstly
the simulation domain is partitioned into several cells, and then each
atom is assigned to these cells by their positions, in the following
the neighbor list is constructed by searching only in neighboring
cells, instead of checking all atom pairs in the system.  It can be
seen that the construction of the neighbor list scales $O(c\cdot N)$,
where $c$ is a constant and independent to $N$. For a system
containing more than 1000 atoms, this approach improves the overall
performance significantly.

For systems with high atomic mobility, we find that the cell edge of
$\frac{1}{2}r_{\text{cut}}$ gives best performance after carrying out
many test runs on some different kinds of computers.

\subsubsection{Partial updating of the neighbor list}

In certain systems where a small number of atoms have high mobility
and most others only oscillate at equilibrium positions, the choice of
skin in Verlet table algorithm can be very tricky. Small value ensures
that atoms which are stored in the list and beyond the cutoff distance
($r_{\text{cut}} < r \le r_{\text{cut}}+r_s $) are as less as
possible, however, updating of neighbor list will become frequent. On
the other hand, the potential/force evaluation efficiency is lowered
when using large skin value, it can be tought to find a good value.

Fortunately, this problem can be solved in the framework of cell
decomposition approach. As searching for neighbors is confined to
several cells, after the neighbor list is out-of-date and needs to be
updated, reconstruction procedure can be carried out for atoms in
relevant cells only, instead of for every atom in the system.

In this method, each cell is accompanied with a ``dirty flag'', which
denotes if the neighbor list corresponding to this cell needs to be
updated. At every timestep all atoms' accumulative displacements will
be checked in each cell, if the sum of the largest two displacements
is greater than $r_s$, then the ``dirty flag'' of this cell and
neighboring cells will be turned on, and the neighbor list
construction procedure is carried out for these cells. After the
construction procedure these flags are turned off.

For aforementioned system this method effectively improves the overall
performance because the CPU time on updating the neighbor list is
minimized. In addition, the skin can be small enough so that the
potential/force evaluation efficiency is maximized.

It needs to be noticed that in this method the neighbor lists of
different atoms should be stored separately (e.g., the neighbor list
array can be defined in the form of \texttt{nlist[N][MAXNL]}, where
\texttt{N} is the number of atoms and \texttt{MAXNL} is the maximum
number of neighbors of an atom). This, however, may slightly increase
the memory requirement although this is not serious in modern
computers where the internal memory is at least hundreds of MB.

\subsubsection{Acceleration of data access by data sorting}

Considering pipeline architectures of modern CPUs, further effort can
be taken to maximize the computing performance: sorting the storage
sequences of atoms in the memory, thus the memory locations of atoms
which in the same cell or neighboring cells are also close to each
other, then the data can be loaded and cached in the CPU more
efficiently.

\begin{figure}[hbtp!]
  \centering
  \includegraphics[width=0.5\textwidth]{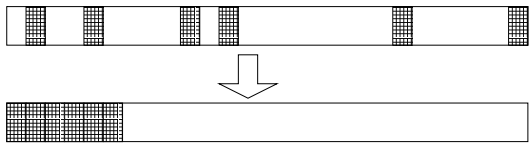}
  \caption{Illustration of data sorting method. In this method, atoms
    which are close to each other in the simulation domain are also
    stored continuously.  Every timestep after updating the neighbor
    list, data of all atoms are sorted to maximize the data cache hit
    rate.}
  \label{fig:data-sorting}
\end{figure}

\begin{figure}[hbtp!]
  \centering
  \includegraphics[width=0.5\textwidth]{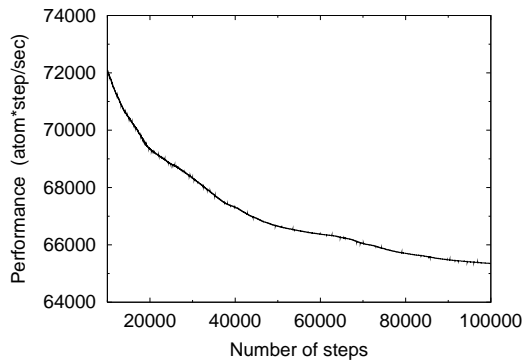}
  \caption{Performance degradation in a gas simulation due to high
    data cache miss rate. The performance is measured in the unit of
    ``atom$\cdot$step/second'', which can be calculated by multiplying
    the number of atoms by the number of steps and divided by the
    elapsed CPU time.}
  \label{fig:cache-pollute}
\end{figure}

For a better understanding of this method, we can consider the
simulation of gas and liquid materials which have high atomic
mobility. In the beginning, after the molecular structure is
initialized, the data of positions, velocities and accelerations are
well ordered and stored in the memory continuously. Sometimes data of
all neighboring atoms can be loaded into CPU data cache if original
data is well organized. But when the simulation is going on and the
atoms are moved here and there, memory locations of these data become
more and more disordered, and the neighboring atoms can be seldom
loaded in the same cache line.  Then the CPU is hard to find the data
of neighboring atoms in the data cache, and has to stall the execution
and fetch required data into the data cache.  However, irrelevant data
along with fresh loaded data pollute the data cache, therefore the
data cache miss rate can be high. This phenomena can be detected in a
long time simulation, as an example shown in
Fig.~\ref{fig:cache-pollute}, where a very significant performance
degradation can be seen.

The solution of this problem is rather straightforward, i.~e., sorting
the data of atoms by their positions and making the memory locations
of the neighboring atoms as close as possible. By adopting this
method, no explicit performance degradation can be seen in long time
simulations. The procedure to carry out the data sorting is shown in
Algorithm \ref{alg:sort}. In this procedure, firstly we find the
longest direction of the domain, and partition the domain into several
layers along this direction, lastly the data of all atoms are stored
layer by layer. Please note that a ``weak'' sorting is used, it means
that data of atoms inside a layer may be disordered. It can be seen
that ``weak'' data sorting procedure scales $O(N)$ as the number of
atoms.

\begin{algorithm}
  \caption{Data sorting procedure in the improved neighbor list algorithm}
  \label{alg:sort}
  \begin{algorithmic}
    \STATE $m \gets$ direction of the longest edge of domain
    \COMMENT{$m=1,2,3$}
    \STATE $nlayer \gets$ number of layers in direction $m$
    \FORALL{$i$ in $1 \cdots N$}
    \STATE $k\gets$ layer index of atom $i$
    \STATE $ numlay(k) \gets  numlay(k)+1$
    \COMMENT{number of atoms in layer $k$}
    \ENDFOR
    \STATE \COMMENT{Find the index of first atom in each layer.}
    \STATE $layer(1) \gets 1$
    \FORALL{$i$ in $1 \cdots nlayer-1$}
    \STATE $layer(i+1) \gets layer(i)+numlay(i)$
    \ENDFOR
    \STATE \COMMENT{Move data to temporary storage and do weak sorting.}
    \FOR{$i=1$ to $N$}
    \STATE $k\gets$ layer index of atom $i$
    \STATE $j \gets layer(k)+numlay(k)-1$
    \STATE $NR(j) \gets R(i), NV(j) \gets V(i)$ 
    \STATE $numlay(k) \gets  numlay(k)-1$
    \ENDFOR
    \STATE $R\gets NR, V\gets NV$
  \end{algorithmic}
\end{algorithm}

Together with the cell decomposition approach in section \ref{sec:cell
  decomp}, the data locality is enhanced.  Thus when parallelizing the
simulation program on SMP platforms, the data can be easily and well
partitioned, and the neighbor list construction can be carried out by
each CPU in the computer simultaneously, thus a high parallelization
execution can be achieved.

In this work, overall procedure for constructing the neighbor
list\footnote{Full neighbor list construction is carried out at the
  first timestep and at every timestep after the data sorting, and
  partial updating is performed between two times of data sorting.} is
shown in Algorithm \ref{alg:bvt}. Here it should be mentioned that
data sorting procedure may not be carried out for low atomic mobility
systems.

\begin{algorithm}
  \caption{The improved neighbor list algorithm}
  \label{alg:bvt}
  \begin{algorithmic}
    \STATE \COMMENT{Assigning each atom into its appropriate cell}
    \FORALL{$i$ in $1\cdots N$}
    \STATE $k\gets$ cell index of atom $i$
    \STATE append $i$ into the list of cell $k$
    \ENDFOR
    \STATE \textbf{call} sorting
    \COMMENT{Sorting atoms by their position. See Algorithm \ref{alg:sort}}
    \FORALL{$i$ in $1\cdots N$}
    \STATE $l \gets$ cell index of atom $i$
    \FORALL{$m \in l$ and neighbors of $l$}
    \FORALL{$j \in$ cell $m$ and $j\ne i$}
    \STATE $\mathbf{r}_{ij}\gets R(j)-R(i)$    
    \STATE apply the periodic boundary condition to $\mathbf{r}_{ij}$
    \IF{$|\mathbf{r}_{ij}| < r_{\text{cut}} + r_s$}
    \STATE append $j$ into the neighbor list of atom $i$
    \ENDIF
    \ENDFOR
    \ENDFOR
    \ENDFOR
  \end{algorithmic}
\end{algorithm}


\section{Results and discussions}
\label{sec:results}

A molecular dynamics simulation program using Lennard-Jones two-body
potential function is developed to compare the performance of three
different neighbor list algorithms. The benchmarks are carried out on
a Compaq Alpha Server DS20 with two EV67/667 MHz processors and Tru64
5.1A operating system installed, and the program is written in Fortran
90 and compiled by Compaq Fortran compiler V5.5--1877. We also run the
same benchmarks on a PC with one Intel Pentium III 866 MHz CPU and Red
Hat Linux 8.0 installed, and a HP RX2600 workstation with two Intel
Itanium2 900 MHz processors and Red Hat Linux Advanced Workstation
installed. Three CPUs have different architectures: Alpha EV67 is a
typical RISC (Reduced Instruction Set Computing) CPU, Intel Pentium
III is a CISC (Complex Instruction Set Computing) CPU, while Itanium2
is declared as a brand new architecture named EPIC (Explicitly
Parallelized Instruction Computing). The performance on three
platforms are different, however the difference among three algorithms
are qualitatively similar.  All results presented in this section are
based on the benchmark on Alpha Server DS20.

In order to measure the performance of an algorithm quantitatively, a
new unit named \emph{atom$\cdot$step/second} is defined. It can be
simply calculated by multiplying the number of atoms and number of
timesteps divided by the number of CPU time elapsed in second.  Larger
value of this quantity stands for better performance. If computer
hardware architectures are identical, and the simulation program
scales $O(N)$ perfectly as the system size, then this quantity is
proportional to the CPU performance.

In our benchmark, firstly some Argon atoms are randomly placed in the
domain according to the predetermined density. Then the molecular
dynamics simulation in canonical ensemble is performed, and the number
of steps is $10^2$ for $10^4$ atoms and more, or $10^3$ for $10^3 \sim
10^4$ atoms, or $10^4$ for 999 atoms or less. The temperature of the
system is 300 K.

The time integration algorithm is implementated by velocity Verlet
scheme, and the canonical ensemble simulation is implemented by using
No\'se--Hoover thermostat \cite{nose}. Real physical units, instead of
reduced units, are used. We collect the parameters in the simulation
and show in Table \ref{tab:param}.

\begin{table}[htbp]
  \centering
  \begin{tabular}{ll}
    \hline
    \textbf{Parameter} & \textbf{Value} \\
    \hline
    $\sigma$        & $3.41$ \AA \\
    $\epsilon$      & $119.8$ kJ/mol \\
    Mass            & $40.0$ atomic unit \\
    Density         & $0.6\; \sigma^{-3}$ \\
    Cutoff distance & $2.5\; \sigma$ \\
    Skin            & $0.5\; \sigma$ \\
    Time step       & $0.8$ fs\\    
    \hline
  \end{tabular}
  \caption{Parameters in the MD simulation}
  \label{tab:param}
\end{table}

The simulation program is parallized using OpenMP technique
\cite{openmp}.  Serial and parallel version of the program are
compiled by turning on or off the relevant compiling options of
OpenMP\footnote{In Compaq Fortran compiler on Tru64/Alpha option
  \texttt{-omp} enables OpenMP, in HP Fortran compiler on
  HP-UX/Itanium option \texttt{+Oopenmp/+Onoopenmp} enables/disables
  OpenMP, and in Intel Fortran compiler option \texttt{-openmp}
  enables OpenMP.}.

In order to verify the improved neighbor list algorithm, all neighbor
lists are dumped to disk files and compared with those in the
execution with Verlet table algorithm. In the verification stage, some
statistical quantities, such as total potential energy, total kinetic
energy, transient temperature of system and the trajectory of all
atoms have been recorded in the interval of 10 timesteps, and data
generated from three algorithms are compared and ensure they differ in
round-off errors only. Intensive tests show that the neighbor lists
from the improved algorithm are exactly the same as those from the
other two algorithms, and three different algorithms give same results
precisely.

Several simulations with different number of atoms and the same
density are carried out, and the performance is calculated and
recorded.

\begin{figure}[htbp!]
  \centering
  \includegraphics[width=0.5\textwidth]{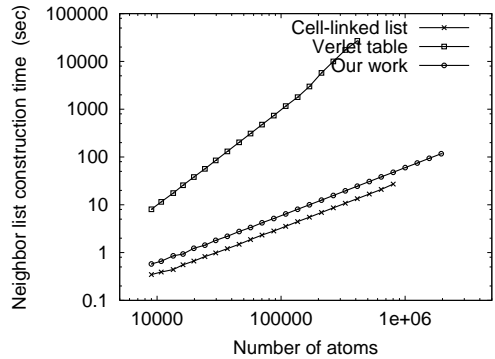}
  \caption{The neighbor list constructing time of three algorithms.
    Three curves from top to bottom denote Verlet table algorithm, the
    improved algorithm and the cell linked list algorithm,
    respectively.}
  \label{fig:nl-construct}
\end{figure}

Firstly we measure the neighbor list constructing time with different
number of atoms for three algorithms, and results are shown in
Fig.~\ref{fig:nl-construct}. In order to measure the CPU time
accurately, a very large number of atoms are used. From the results we
can see that the neighbor list constructing time of Verlet table
algorithm scales $O(N^2)$ as the number of atoms, while the other two
algorithms scale $O(N)$.  We can also see that the neighbor list
constructing time of the improved algorithm is slightly larger than
that of the cell linked list algorithm. However, the neighbor list
needs to be constructed in every timestep in the cell linked list
algorithm, but it needs not in our improved algorithm, thus the CPU
time used to maintain the neighbor list in the improved algorithm is
much less than those in the other two algorithms.

\begin{figure}[hbtp!]
  \centering
  \includegraphics[width=0.5\textwidth]{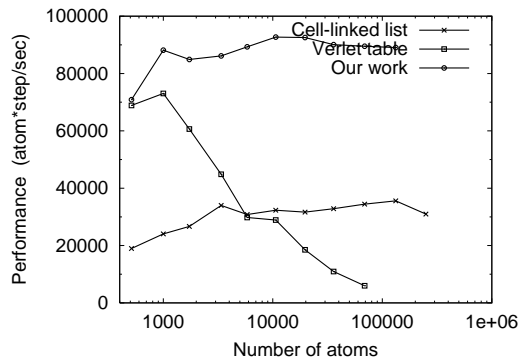}
  \caption{Comparison of three algorithms on a single processor system.
    The performance is measured in the unit of
    ``atom$\cdot$step/second'', which can be calculated by multiplying
    the number of atoms by the number of steps and divided by the
    elapsed CPU time. The three curves from top to bottom denote the
    performance of the improved algorithm, Verlet table algorithm and
    the cell linked list algorithm, respectively.}
  \label{fig:comp}
\end{figure}

\begin{figure}[hbtp!]
  \centering
  \includegraphics[width=0.5\textwidth]{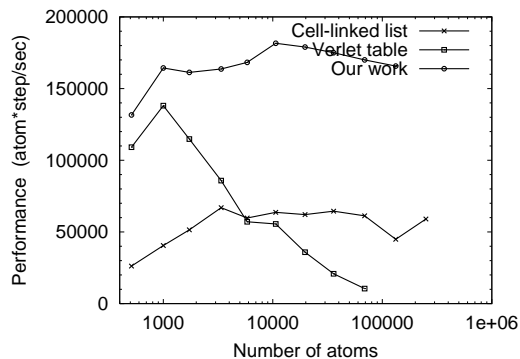}
  \caption{Comparison of three algorithms on a dual-processor system. The 
    three curves from top to bottom denote the performance of the
    improved algorithm, Verlet table algorithm and the cell linked
    list algorithm, respectively. }
  \label{fig:comp-smp}
\end{figure}

Then we carry out the real benchmark and calculate the overall
performance of the simulation with different algorithms. Results are
shown in Fig.~\ref{fig:comp} (single processor results) and
Fig.~\ref{fig:comp-smp} (dual-processor results)

From the results of Fig.~\ref{fig:comp}, we can know that the proposed
improved algorithm improves the overall performance of the molecular
dynamics simulation significantly.  When the system is small, the
performance of the improved algorithm is as high as that of Verlet
table algorithm. As the system size increases and there are more and
more atoms, the performance of Verlet table algorithm decreases
rapidly, but the performance of the improved algorithm becomes even
better. As the system is as big as $7 \times 10^5$ atoms, the
performance of Verlet table algorithm becomes very bad, and is much
lower than that of the improved algorithm.  On the other hand,
although the cell linked list algorithm can handle a very large system
due to its small memory requirement, its performance is much lower
than that of the improved algorithm.  For large systems, the improved
algorithm in our work is about 2 $\sim$ 3 times faster than
conventional neighbor list algorithms.

From Fig.~\ref{fig:comp-smp} we can see that the improved algorithm in
SMP platforms exhibits higher performance than the other two
algorithms.  The conclusions in above discussions for single-processor
are also valid for dual-processor.


\section{Conclusions}
\label{sec:con}

Nowadays as the continuously increasing of computing power, huge and
complicated molecular simulations will be attempted and involve a
larger number of atoms and more complex potential functions.  The
expectation of running molecular simulations faster and easier for
larger systems on existing platforms makes it important to improve the
neighbor list algorithm in order to reduce the unnecessary interatomic
distance calculations. In this paper, we proposed an improved order
$O(N)$ neighbor list algorithm which incorporates the advantages of
conventional Verlet table and cell linked list algorithms. In the new
algorithm, cell decomposition approach is adopted to accelerate the
neighbor list construction speed, and data sorting method is used to
maximize the CPU data cache hit rate. In addition, partial updating of
the neighbor list is introduced to minimize unnecessary neighbor list
reconstructions. We carried out the benchmarks to these algorithms
using molecular dynamics simulations with Lennard-Jones potential
function, and compared the performance of the improved algorithm with
that of the other two algorithms. The results show that the improved
algorithm outperforms conventional Verlet table and the cell linked
list algorithms by a factor of $2 \sim 3$ in single-processor and SMP
platforms.


\section*{Acknowledgments}
\label{sec:ack}

This work is supported by the Singapore-MIT Alliance.

\end{document}